\documentclass[usenatbib]{mn2e}
\usepackage{mathtools}
\usepackage{graphicx}
\usepackage{epstopdf} 
\usepackage{natbib}
\usepackage{verbatim}

 \usepackage{times}
\def\lsim{\mathrel{\rlap{\lower4pt\hbox{\hskip1pt$\sim$}}
    \raise1pt\hbox{$<$}}}                
\def\gsim{\mathrel{\rlap{\lower4pt\hbox{\hskip1pt$\sim$}}
    \raise1pt\hbox{$>$}}}                

\begin{document}

\title[]
{Spectroscopic detections of CIII]$\lambda$1909 \AA\ at $z\simeq6-7$: A  new probe of early star forming galaxies and cosmic reionisation}

\author[Stark et al.] 
{Daniel P. Stark$^{1}$\thanks{dpstark@email.arizona.edu}, 
Johan Richard$^2$, 
St\'{e}phane Charlot$^3$, 
Benjamin Cl\'{e}ment$^{1,2},$ \newauthor
Richard Ellis$^{4}$,
Brian Siana$^{5}$,
Brant Robertson$^{1}$,
Matthew Schenker$^{4}$, 
Julia Gutkin$^3$, \newauthor
\& Aida Wofford$^3$  \\
$^{1}$ Steward Observatory, University of Arizona, 933 N Cherry Ave, Tucson, AZ 85721 USA \\  
$^{2}$ Centre de Recherche Astrophysique de Lyon, Universite Lyon 1, 9 Avenue Charles Andre, 69561,  France  \\
$^{3}$ UPMC-CNRS, UMR7095, Institut dÕAstrophysique de Paris, F-75014 Paris, France \\
$^{4}$ Cahill Center for Astronomy \& Astrophysics, California Institute of Technology, Pasadena, CA 91105 USA \\
$^{5}$ Department of Physics \& Astronomy, University of California, Riverside, CA 92507 USA \\
}
\date{Accepted ... ;  Received ... ; in original form ...}

\pagerange{\pageref{firstpage}--\pageref{lastpage}} \pubyear{2014}

\hsize=6truein
\maketitle

\label{firstpage}
\begin{abstract}

Deep spectroscopic observations of $z\gsim 6.5$ galaxies have revealed a marked decline with increasing
redshift in the detectability of Ly$\alpha$ emission. While this may offer valuable insight into the end of the reionisation
process, it presents a fundamental challenge to the detailed spectroscopic study of the many hundreds of
photometrically-selected distant sources now being found via deep HST imaging, and particularly those
bright sources viewed through foreground lensing clusters. In this paper we demonstrate the validity of a new
way forward via the convincing detection of an alternative diagnostic line, CIII]$\lambda$1909 \AA\ , seen in 
spectroscopic exposures of two star forming galaxies at $z_{\rm{Ly\alpha}}=6.029$ and 7.213. The former detection is based on 
a 3.5 hour X-shooter spectrum of a bright  ($J_{\rm 125}=25.2$) gravitationally-lensed galaxy behind the cluster Abell 383.
The latter detection is based on a 4.2 hour MOSFIRE spectra of one of the most distant spectroscopically confirmed galaxies,
GN-108036, with $J_{\rm{140}}=25.2$. Both targets were chosen for their continuum brightness and previously-known
redshift (based on Ly$\alpha$), ensuring that any CIII] emission would be located in a favorable portion of the
near-infrared sky spectrum. We compare our CIII] and Ly$\alpha$ equivalent widths in the context
of those found at $z\simeq 2$ from earlier work and discuss the motivation for using lines other than
Ly$\alpha$ to study galaxies in the reionisation era. 

\end{abstract}


\section{Introduction}

In the last few years, HST imaging has delivered several hundred 
galaxies with photometric redshifts estimated to be above $z\simeq
7$ (e.g., \citealt{Schenker2013,McLure2013,Bouwens2014,Ellis2013}).   Yet concerted
efforts to secure spectroscopic redshifts through the detection of
Ly$\alpha$  emission have led to minimal progress.  At the time of
writing there are fewer than 10 spectroscopically-confirmed galaxies beyond
$z\simeq 7$  \citep{Vanzella2011,Ono2012,Schenker2012, Finkelstein2013,Schenker2014}.
and no convincing example beyond $z=7.62$.   
While this represents some progress compared to 2006 when the frontier was at redshift 6.96
\citep{Iye2006}, the investment of telescope time to push the frontier back a mere 100 Myr
in look-back time has been very costly.  The relevant spectroscopic campaigns probing
beyond $z\simeq 7$ have achieved a success rate for Ly$\alpha$ significantly lower than at 
$z\simeq 5-6$, where large equivalent width Ly$\alpha$ emission is found in more 
than half of the galaxies observed (Stark et al. 2010, 2011). The most natural explanation 
for the downturn in Ly$\alpha$ transmission in similarly-selected star-forming galaxies
is that the intergalactic medium (IGM) is still partially neutral at $z\simeq 7$, causing 
attenuation of  Ly$\alpha$ (e.g., \citealt{Treu2013,Schenker2014}).

\begin{figure*}
\begin{center}
\includegraphics[trim=0.15cm 0cm 1.0cm 0cm,width=1.0\textwidth]{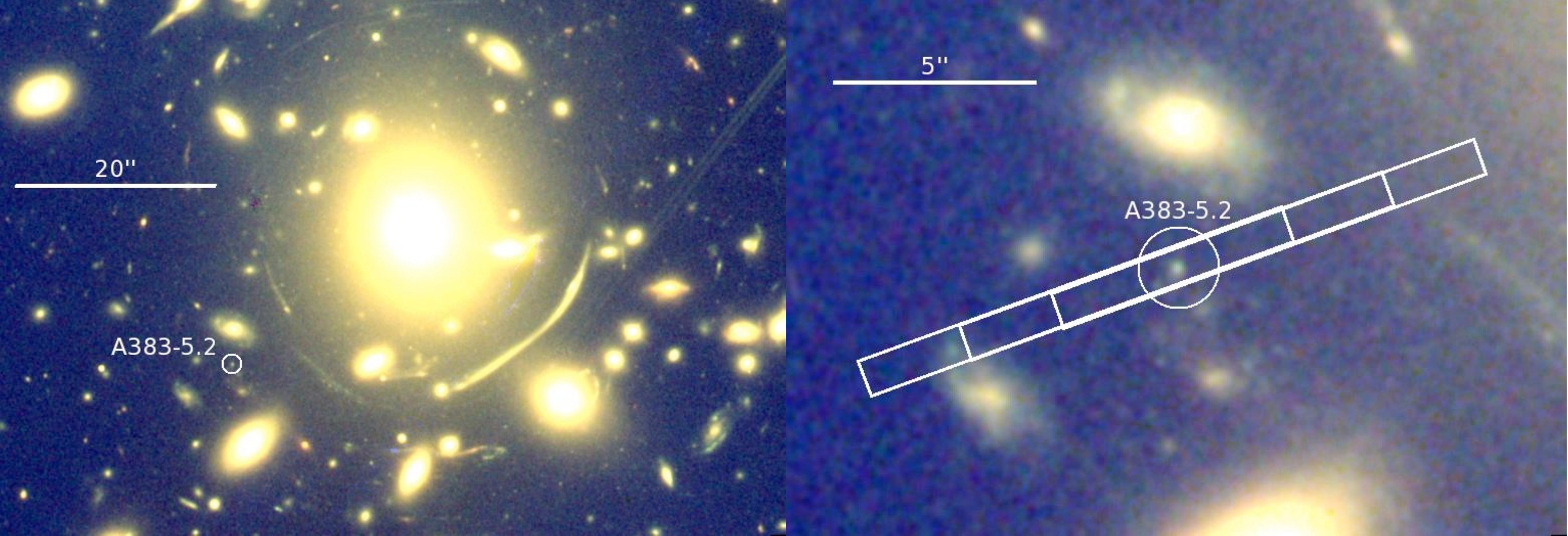}
\caption{Overview of the VLT/XShooter observations of the $\rm{z_{Ly\alpha}=6.027}$ galaxy A383-5.2 first reported in 
Richard et al. (2011). 
({\it Left:}) location of the targeted image A383-5.2 with respect to the cluster centre. 
({\it Right:})  Position and orientation of the XShooter slit, showing the three dither positions.
}
\label{fig:hst}
\end{center}
\end{figure*}

While the reduced transmission of Ly$\alpha$ at $z\gsim 7$ provides valuable insight
into the end of the reionisation process, it also has profound implications for future 
spectroscopic studies of galaxies within the reionisation era. The arrival
of efficient near-infrared spectrographs \citep{McLean2012} and the discovery of a 
growing number of  bright gravitationally-lensed galaxies (e.g., \citealt{Richard2011,Zitrin2012,Bradley2013}) 
promises great progress not only in measuring
redshifts but also in spectroscopic diagnostic studies of typical early galaxies.  
However, with Ly$\alpha$ increasingly obscured, presumably by the partially-neutral IGM at $z>7$, 
the current approach of targeting  Ly$\alpha$ is unlikely to be productive.  Indeed, even
a 52 hour exposure recently undertaken with the ESO VLT failed to detect the line
in a promising bright $z\simeq 7$ target \citep{Vanzella2014}. 

The disappointing outcome of earlier spectroscopic efforts raises several important
strategic questions.  If Ly$\alpha$ is undetectable in most star-forming galaxies in the reionisation
era, must we rely entirely on photometric redshifts and imaging data to determine
their physical properties? In fact, if there are no detectable features in the rest-frame UV, 
it would be unclear whether ground-based telescopes, including more powerful ones
soon under construction, can contribute much to the study of early galaxies. 
Moreover, while JWST will ultimately provide access to strong nebular emission lines ([OIII]$\lambda$5007, H$\alpha$) 
to $z\sim8$, spectroscopic studies at the highest redshifts  ($z > 11-15$), where new 
sources are likely to be found, must perforce rely on detectable rest-frame UV features.  

Ly$\alpha$ is usually considered to be the only prominent UV emission line for star-forming
galaxies.  But this  perception is mostly based on spectroscopic studies of relatively massive, 
chemically-enriched galaxies undergoing fairly rapid ($\gsim 10$ M$_\odot$ yr$^{-1}$) star formation 
(e.g., \citealt{Shapley2003}).  However, the galaxies that populate the reionisation era are likely to
have lower masses, reduced metallicities and larger specific star formation rates.  In an important,
in-depth, study of a metal-poor (1/6 Z$_\odot$) galaxy at $z\sim 2$, \citet{Erb2010} demonstrated   
the presence of numerous prominent emission lines (OIII]$\lambda\lambda$1661,1666; He II$\lambda$1640, 
and the blended CIII]$\lambda$1908 \AA\ doublet) throughout the rest-frame UV spectral region.
Additional low mass galaxies have recently shown similar rest-UV emission spectra  
\citep{Christensen2012,Bayliss2013,James2014}.

In an attempt to select galaxies with physical properties similar to those expected in the reionisation era,
Stark et al. (2014) secured deep rest-frame UV spectra for 17 faint gravitationally-lensed galaxies at 
$z\simeq 1.5-3$ with characteristic very blue UV colors ($<\beta>\simeq-2.2$) and intrinsic (unlensed)
stellar masses ranging from 2$\times$10$^6$  M$_{\odot}$ to 1.4$\times$10$^9$ M$_{\odot}$. In this campaign, the  CIII]$\lambda$1908 
\AA\ doublet was typically the strongest line other than Ly$\alpha$ and detected in 16 galaxies with 
rest-frame equivalent widths up to 14~\AA.   Stark et al. interpreted such powerful metal line emission 
as arising from the large ionisation parameters and electron temperatures associated with metal poor galaxies 
dominated by very young stellar populations.     

Thus, based on the emerging picture at $z\simeq 2$,  the strongest UV metal lines  (CIV$\lambda$1549, 
OIII]$\lambda\lambda$1661,1666, CIII]$\lambda$1908 \AA\ ) provide the most promising route not only
to securing redshifts of star-forming galaxies in the reionisation era but also to progress in determining
their physical nature. Stark et al. (2014) predicted that existing ground-based facilities should be   
able to detect these UV metal emission lines in $z\gsim 6$ galaxies. In this paper we present the first 
confirmation of this prediction from an ongoing campaign targeting UV metal line emission in the spectra of $z\gsim 6$ galaxies.  
In the initial phase of this program, our primary goal has been to verify that metal line equivalent widths are 
sufficiently large to be detected at $z\gsim 6$.  To maximise the likelihood of success, we have 
chosen targets that have bright continuum magnitudes  ($\rm{H}\lsim 25.5$) and known (Ly$\alpha$-based) 
spectroscopic redshifts.  The latter requirement ensures that  the relevant metal emission lines would lie in regions where the 
atmospheric transmission is near unity and obscuration from sky emission lines minimal.  Clearly in
the longer term, it is desirable to demonstrate spectroscopic detections for systems where Ly$\alpha$
is completely suppressed.

Here we report VLT/X-Shooter and Keck/MOSFIRE spectroscopic observations of the CIII]$\lambda$1908 
doublet in two $z>6$ galaxies.   The first system we discuss is a bright gravitationally-lensed system at $z=6.03$.   
The galaxy,  reported in Richard et al. (2011) and Bradley et al. (2013), is multiply-imaged into a pair 
by the foreground cluster Abell 383.  The two images, A383-5.1 and A383-5.2, are both bright (J$_{125}$=24.6 and 25.2) and highly-magnified 
(11.4$\times$ and 7.3$\times$).  Ly$\alpha$ emission was detected in each image by Richard et al. (2011) and,
at the redshift $z_{\rm{Ly\alpha}}=6.029$, the CIII] doublet  falls within a relatively clean region of the $J$-band.  The second system we discuss is 
GN-108036, a $z=7.2$ galaxy located in the GOODS North field.    This galaxy was first identified as a  bright 
(J$_{140}$=25.2) z-band dropout in  \citet{Ouchi2009}.  Ly$\alpha$ emission was reported in Ono et al. (2012), 
providing spectroscopic confirmation at $z_{\rm{Ly\alpha}}=7.213$.   At this redshift, the CIII] doublet is located in a region free 
from atmospheric absorption within the H-band.

\begin{figure*}
\begin{center}
\includegraphics[width=1.0\textwidth]{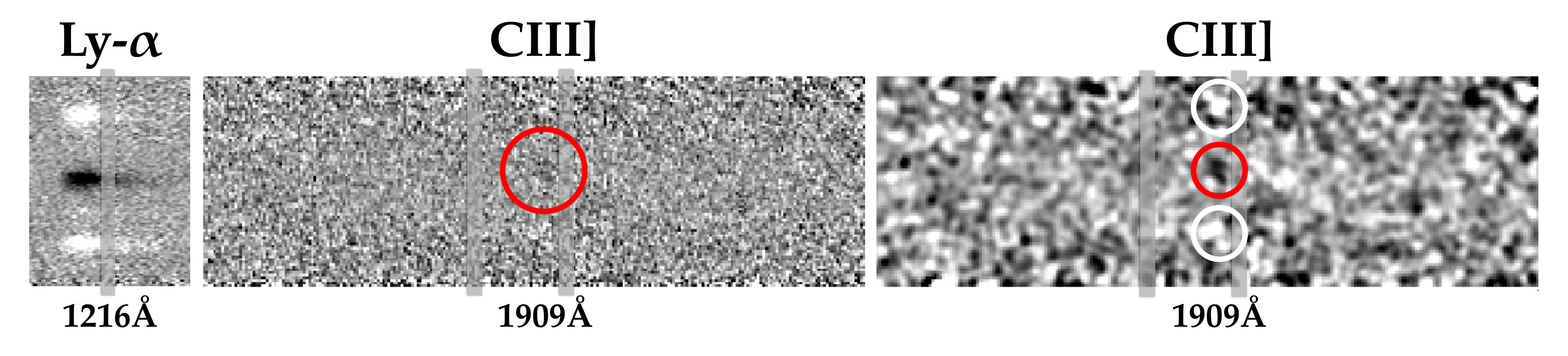}
\caption{XShooter 2D spectrum of the $\rm{z_{Ly\alpha}=6.027}$ galaxy A383-5.2.   The leftmost panel is 
centred on Ly$\alpha$ emission which is detected with the visible arm at 0.8546 $\mu$m.  The 
location of CIII]$\lambda$1909 at 1.3412$\mu$m is shown in the middle (unsmoothed) and 
rightmost (smoothed) panels.   The [CIII]$\lambda$1907 emission line is under the OH sky line 
just blueward of CIII]$\lambda$1909.        }
\label{fig:spec}
\end{center}
\end{figure*}

The paper is organised as follows.  We discuss our new near-infrared spectroscopic observations and reduction 
procedures in \S2.  The properties of Ly$\alpha$ and CIII] for both sources are detailed in \S3.   We describe the 
modelling of the continuum and emission lines for A383-5.2 in \S4.   By determining the velocity offset of Ly$\alpha$
with respect to CIII] we discuss the implications for the interpretation of the evolution of the fraction of photometrically-selected 
galaxies with Ly$\alpha$ emission  in \S5. Finally,
we discuss the prospects of using CIII] in future surveys and summarise our conclusions in \S6.
Throughout the paper, we adopt a $\Lambda$-dominated, flat Universe
with $\Omega_{\Lambda}=0.7$, $\Omega_{M}=0.3$ and
$\rm{H_{0}}=70\,\rm{h_{70}}~{\rm km\,s}^{-1}\,{\rm Mpc}^{-1}$. All
magnitudes in this paper are quoted in the AB system \citep{Oke1983}.

\section{Observations and data reduction}

\subsection{VLT/XShooter}
As part of  ESO program ID: 092.A-0630 (PI: Richard), the galaxy A383-5.2 \citep{Richard2011} was observed with the 
XShooter spectrograph on the VLT \citep{Vernet2011}.    We chose to pursue A383-5.2 because A383-5.1 (the brighter 
of the two lensed images)  is located closer to the cluster centre, where diffuse light from bright cluster galaxies contributes 
considerably to the background in the near-infrared.  Observations were conducted on the nights of 2013 October 27 
and 2013 December 15 for 3 and 2 observing blocks (OBs) of 1 hour each, respectively.  We used the 
11$\times$0.9" slit oriented to avoid bright  galaxies (Figure 1).   One OB comprised three exposures of 
955 sec. in the visible arm, covering the wavelength range 5630-10090 \AA\ at a resolution of R$\sim$8800, 
and 3 exposures of 968 sec. in the near-infrared  
arm (with 4 sub-integrations of 242 sec.), covering the wavelengh range 10350-24780 \AA\ at a resolution of 
R$\sim$5300.   A dither pattern of $\pm$2.5" along the slit was performed between each exposure, for optimal sky subtraction 
(Figure 1, right). The total exposure time on source was 14325 sec in the visible arm and 14520 sec in the near-infrared arm.  

The sky conditions were clear and the seeing was very good in the first 3 OBs, with a range of 0.50-0.70" and a median seeing of 0.55", 
but less good in the second set of 2 OBs, with a range of 0.60-0.90". 
A spectroscopic standard star was observed on both nights for absolute flux calibration, and multiple telluric 
standard stars were observed to estimate telluric correction.

We used the latest version of the XShooter data reduction software (v.2.2.0) in the Recipe Flexible Execution Workbench 
(REFLEX) environment to perform 
a first calibration and reduction of each exposure. We then applied standard IDL and IRAF routines for 
optimally combining and extracting the 15 exposures. Specifically, we used the Lyman-$\alpha$ emission line, 
well-detected in each exposure, to correct for variations in seeing and atmospheric conditions between 
the different OBs, and applied a scaling and weighting of the 2D spectra according to the flux and detection 
level of the Lyman-alpha line. We also used the spatial position of the line measured in the reduced spectrum to 
precisely compute the offsets between each OB for optimal combination. We used the same offsets, scaling factors and 
weights to combine the exposures in the near-infrared arm.  Applying these corrections slightly 
strengthened the S/N of the CIII] line, increasing our confidence in the detection.

The combined 2D spectrum was extracted using a variance-weighting scheme in IRAF based on the detected 
profile of the lines. We also used the normalized extracted spectrum of the telluric standards to apply a median 
correction for telluric absorptions in the near-infrared arm. The final combined 2D  spectrum at wavelengths near  
Ly$\alpha$ and CIII] are presented in Figure 2 and discussed in Section 3.1

\subsection{Keck/MOSFIRE}

We secured spectroscopic observations of GN-108036 with MOSFIRE \citep{McLean2012} on the Keck I telescope on the 
nights of March 6 and April 11, 2014. GN-108036 is one of the most distant spectroscopically-confirmed 
galaxies at $z$=7.213, first verified by Ono et al (2012) based on three separate exposures with the 
 DEep Imaging Multi-Object Spectrograph (DEIMOS)  on Keck 2; the rest-frame equivalent width of Ly$\alpha$ is reported as 33 \AA\ .
During the March run, we compiled a total of 3.1 hours of exposure in the H band with an 0.8" slit. 
Conditions were generally clear, but with slight cloud during the first half of the observations. The median 
FWHM of a reference star included on the mask was 0.6". On the April night, we secured an 
 additional 1.1 hours, with a median seeing FWHM of 0.5" and clear conditions.
 
The data were reduced using the regular MOSFIRE data reduction pipeline (DRP)\footnote{\rm https://code.google.com/p/mosfire/} 
following a procedure similar to that described in \citep{Schenker2014} to which the reader is referred. Briefly,
this pipeline first creates a median, cosmic-ray subtracted flat field image for each mask. Wavelength solutions for
each slit are fit interactively for the central pixel in each slit, then propagated outwards to the slit edges to derive a full wavelength grid. 
Background subtraction is handled as a two stage procedure. First, individual stacks of all A frames and all B frames are used to 
produce A-B and B-A stacks. As the A and B frames are temporally interleaved, this provides a first level of subtraction. 
Secondly, a 2-D b-spline model is fit to the residuals in each of these stacks. The two stacks are then shifted, rectified, 
and combined, producing a positive source signal flanked by two negative signals at approximately half strength, separated 
by the dither length. To account for the variation in conditions during the March run, we split the data into three segments 
of $\simeq$1 hour each,  which were then run through the DRP separately. To produce a final science stack, the resulting 
three images plus the single reduced image from our April data were then stacked using inverse-variance weighting. 
The final stack has an average 5$\sigma$ flux limit of 1.8 $\times$ 10$^{-18}$ erg cm$^{-2}$ s$^{-1}$ in between sky lines, consistent with the 
expectation produced from the MOSFIRE exposure time calculator. The final combined 2D spectrum at a wavelength
near CIII] is presented in Figure 3 and discussed in Section 3.2 below.

\section{Rest-UV Spectroscopic Properties}
\subsection{A383-5.2}
\subsubsection{Large equivalent width Ly$\alpha$ emission}

The Ly$\alpha$ emission line of both A383-5.1 and A383-5.2 was first reported in the original \citet{Richard2011}
paper following observations with the DEIMOS spectrograph on the Keck II telescope.    
In the discovery spectrum, Ly$\alpha$ was significantly detected (S/N=7.3), but it was blended with a sky line and the slit was 
slightly offset from the target (J. Richard, private communication), precluding reliable constraints on the 
line strength and profile.   

Figure 2 demonstrates a much more significant detection of  Ly$\alpha$.  The improved spectral resolution of XShooter 
separates Ly$\alpha$ from the nearby sky line.  We derive a total Ly$\alpha$ line flux of 1.2$\times$10$^{-16}$ erg cm$^{-2}$ s$^{-1}$.   
Correcting for slit losses (taking into account  the size of the galaxy, the slit width, and the seeing), we estimate an aperture 
correction of 1.06.   The peak of the line occurs at 8545.6~\AA, implying a Ly$\alpha$ redshift of z$_{\rm{Ly\alpha}}=6.0294$.   
After correcting for the source magnification (7.3$\times$; Richard et al. 2011) and applying the aperture correction, the total 
Ly$\alpha$ luminosity is 6.8$\times$10$^{42}$ erg s$^{-1}$.   

The Ly$\alpha$ equivalent width is inferred via two different methods.  First, using the measured line flux and the 
upper limit (2$\sigma$) on the continuum flux from the XShooter spectrum ($<4.8\times10^{-19}$ erg cm$^{-2}$ 
s$^{-1}$ \AA$^{-1}$), we derive a lower limit on the rest-frame equivalent width (W$_{\rm{Ly\alpha,0}}> 83$~\AA).    
Since the continuum is usually well below spectroscopic flux limits, it is common to use the continuum 
flux measured from broadband imaging to calculate the equivalent widths of emission lines.   With this second method, 
we infer W$_{\rm{Ly\alpha,0}}$=138~\AA, consistent with the upper limit derived above and 
placing A383-5.2 among the most  extreme  Ly$\alpha$ emitters at $z\simeq 6$ 
(e.g., \citealt{Ouchi2010,Stark2011,Pentericci2011,Curtislake2012}).
The large equivalent width likely reflects both a larger-than-average escape fraction of 
Ly$\alpha$ radiation (f$_{\rm{esc,Ly\alpha}}$) and  an efficient production rate of Ly$\alpha$ photons
(as might be expected for a young metal-poor system).   
Assuming a metallicity of 0.2 Z$_\odot$, a Chabrier IMF, and 10\% ionising photon escape fraction) and 
using \citet{Bruzual2003} stellar population models, the Ly$\alpha$ luminosity of A383-5.2 implies a 
a star formation rate of 3.2 (f$_{\rm{esc,Ly\alpha}}$)$^{-1}$ 
M$_\odot$ yr$^{-1}$.   

\subsubsection{CIII]$\lambda$1909 emission}

At $z\simeq 2-3$, the CIII] equivalent width generally increases with the Ly$\alpha$ equivalent width (e.g., \citealt{Shapley2003}, 
Stark et al. 2014).   For Ly$\alpha$ emitters with equivalent widths as large as A383-5.2 (W$_{\rm{Ly\alpha,0}}$=138~\AA), the  
blended CIII]$\lambda$1908 doublet would be expected to have a rest-frame equivalent width above 10~\AA.  However at 
$z\gsim 6$ the resolution of XShooter resolves the CIII] doublet, so each of the individual components will be detected with somewhat 
lower equivalent widths. Given the bright apparent magnitude of A383-5.2 (J$_{\rm{125}}$=25.2),  both components of 
CIII] should be above the XShooter detection limits if located in between the OH sky lines.

Knowledge of the spectroscopic redshift from Ly$\alpha$ allows us to estimate the observed wavelengths of  CIII].    
However, a precise determination must account for the fact that the peak of the emergent Ly$\alpha$ profile is 
typically offset with respect to the systemic redshift probed by CIII].  For low mass star forming galaxies similar to A383-5.2, 
the centroid of the blended CIII] doublet is blueshifted between 60 and 450 km s$^{-1}$ (with a mean of 
320 km s$^{-1}$) from Ly$\alpha$ (Stark et al. 2014).   We thus expect [CIII]$\lambda$1907 to lie between 1.3383 and 1.3400 $\mu$m 
and CIII]$\lambda$1909 to lie between 1.3397 and 1.3415 $\mu$m, each of these windows spanning just 50 spectral pixels.  
Since the spatial position is also well established (by the location of Ly$\alpha$ along the slit), the window over which 
we expect to see CIII] is confined to a very small area in the  spectrum.   

Examination of the XShooter spectrum reveals a 3.8$\sigma$ emission feature  at 1.3412 $\mu$m, in the spectral window 
expected for the CIII]$\lambda$1909 emission line.  The feature is in the centre of the slit at the spatial location of A383-5.2 
and is blueshifted by 120 km s$^{-1}$ with respect to Ly$\alpha$, implying a CIII]$\lambda$1909 redshift of $z=6.0265$.  
At this redshift, [CIII]$\lambda$1907 is unfortunately coincident with an atmospheric OH  line at 1.3397$\mu$m.  There 
is a positive emission feature at the expected location, but the noise associated with the sky line precludes useful flux constraints.
The  flux of CIII]$\lambda$1909 is $\rm{2.1\pm 0.6\times10^{-18}~erg~cm^{-2}~s^{-1}}$.    As we will show below, 
the chance of randomly finding a 3.8$\sigma$ fluctuation in the small 50 pixel area defined by Ly$\alpha$ is very small.

To compare the UV spectrum of A383-5.2 to that of UV line emitters at $z\simeq 2-3$ (e.g. \citealt{Erb2010,James2014}, 
Stark et al. 2014), we must estimate the flux in the [CIII]$\lambda$1907 line.   The ratio of the individual 
components of CIII] is set by the electron density.  Over the range of gas densities typical in high redshift star forming galaxies 
(10$^{2}$  cm$^{-3}$ and 10$^{4}$ cm$^{-3}$; e.g., \citealt{Hainline2009,James2014}), the [CIII]$\lambda$1907/CIII]$\lambda$1909 ratio varies between 1.6 and 1.2
respectively.   Taking the ratio as 1.4$\pm$0.2, we estimate that the total flux in the CIII] doublet is  $\rm{5.4\pm 1.4 \times10^{-18}~erg~cm^{-2}~s^{-1}}$.  
Via the same procedure, we determine the total rest-frame CIII] equivalent width to be $12.7\pm3.5$~\AA.   

Confidence in the detection is bolstered by two tests.  As described above, prior knowledge of the redshift allows us to define 
a very small (50 pixels by 5 pixels) search window for each component of the CIII] doublet.   The likelihood 
of randomly finding a 3.8$\sigma$ emission feature in the  area expected for CIII] is highly unlikely.  In N=10$^5$ realisations 
of the error spectrum, we do not find any case in which the flux in the small window is as large as we observed.    We  have also verified 
that the emission feature is not dominated by spurious emission in one or two of the fifteen exposures.   We visually 
examined each of the exposures and verified that hot pixels and cosmic rays are not present.   We also split the 
exposures into two subsets.   The first group contains the six exposures with the best seeing and stable conditions 
(as defined by the Ly$\alpha$ detection in the visible arm), and the second group contains the nine frames over which 
the seeing degraded.   The CIII] emission feature is 2.5$\times$ more significant in the 
stack with the best seeing, as would be expected if the line is associated with A383-5.2 and not a detector artifact. 

\subsection{GN-108036}
\begin{figure}
\begin{center}
\includegraphics[width=0.5\textwidth]{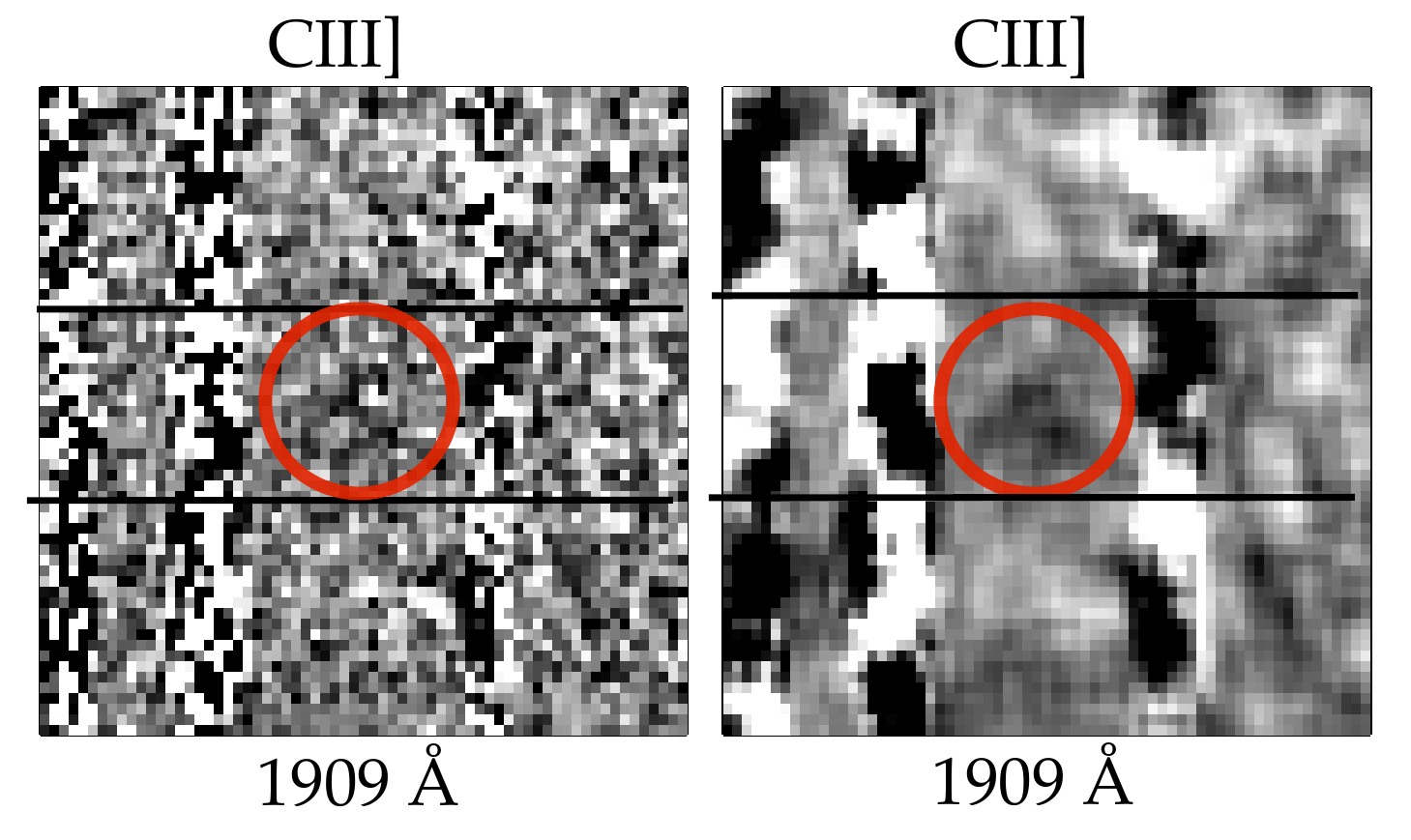}
\caption{MOSFIRE 2D H-band spectrum of the $z=7.213$ galaxy GN-108036 (unsmoothed in left panel, smoothed in right  panel).  
The spectroscopic redshift from Ono et al. (2012) allows the spectrum to be converted to the rest-frame.   Black horizontal lines define the 
location of the galaxy on the MOSFIRE slit.    The red circle denotes the expected wavelength and spatial position 
of the CIII]$\lambda$1909 emission line.   The  flanking negative signals resulting from the A-B dither pattern and 
subtraction are seen above and below the red circle.}
\label{fig:spec}
\end{center}
\end{figure}

The centroid of Ly$\alpha$ (0.9984~$\mu$m) reported for GN-108036 by Ono et al. (2012) places CIII] in the H-band.  
Assuming the velocity offset of Ly$\alpha$ with respect to CIII] is between 0 and 450 km s$^{-1}$ (e.g. Stark et al. 2014), [CIII]$\lambda$1907 
will be located between 1.56359 and 1.56594 $\mu$m and CIII]$\lambda$1909 between 1.56528 and 1.56763$\mu$m.    
With the spectral resolution of MOSFIRE, each window  spans only 14 spectral pixels.  Figure 3 reveals a faint emission feature in the 
small window as expected for CIII]$\lambda$1909.  Clearly the feature is not as prominent as in A383-5.2 (Figure 2) but this is 
perhaps not surprising given the lower Ly$\alpha$ equivalent width (33~\AA).    The line is in a region free from skyline contamination, and 
the centroid is consistent with the expected position of the target along the slit to within 3 pixels, or 0.55 arcseconds.   
We measure a flux of 0.9$\pm0.3\times$10$^{-18}$ erg cm$^{-2}$ s$^{-1}$ 
in an aperture of 1.1 arcseconds in the spatial direction and 13 \AA\ in the spectral direction.   
With a S/N of 2.8, it is difficult to reliably determine the line centroid.   Positive emission roughly 
spans 1.5672 to 1.5684$\mu$m.   If the emission shown in Figure 3 is CIII]$\lambda$1909, it would imply a small velocity offset 
for Ly$\alpha$ (between $-$150 and 76 km s$^{-1}$), consistent with Ly$\alpha$ emerging near the systemic redshift.
At the redshift of GN-108036, the two components of the CIII] doublet are split by 17~\AA.   For the CIII]$\lambda$1909 
redshift defined by the emission feature described above, the [CIII]$\lambda$1907 line would lie underneath the 
bright OH skyline visible blueward of the CIII]$\lambda$1909 in Figure 3.    The noise is more than 3$\times$ larger on the 
skyline, precluding identification of the second component.

One test of the reality of the detection is to measure the fluxes at the expected 
positions of the two negative images flanking the emission feature. As our final stack is the sum of an A-B image and a shifted 
B-A image,  we expect two negative images, each of approximately half the total flux, offset above and below the positive image 
by the dither length of 2.5 arcseconds. We find this is the case and measure significances of 2.0$\sigma$ above the line, and 1.9$\sigma$ 
below. We further evaluate the line in the four separate subsets of the data discussed earlier, each with an exposure time of $\simeq$1 hour. 
In each subset, we measure a positive signal to noise ratio in the same aperture used to measure the composite line flux.

To compute the total equivalent width of the CIII] doublet, we follow same procedure we described in \S3.1.2.  Given the likely 
range of electron densities (10$^{2}$ cm$^{-3}$ - 10$^{4}$  cm$^{-3}$), we expect a [CIII]$\lambda$1907/CIII]$\lambda$1909 
flux ratio of 1.4$\pm$0.2.   Taking into account the uncertainty in the flux ratio and the noise in the CIII]$\lambda$1909 flux, we 
estimate a   [CIII]$\lambda$1907 flux of 1.3$\pm0.5\times$10$^{-18}$ erg cm$^{-2}$ s$^{-1}$.   The CIII] doublet 
rest-frame equivalent width implied by this analysis ($7.6\pm 2.8$~\AA) is slightly lower than seen in A383-5.2, but is consistent with 
the Ly$\alpha$ - CIII] equivalent width relationship for $z\simeq 2-3$ galaxies from Stark et al. (2014).

Although our detection of CIII]$\lambda$1909 is much less secure than that of A353-5.2, it is a good indication of the
challenge and prospects of securing non-Ly$\alpha$ redshifts for typical $z>7$ sources. Deeper MOSFIRE data with exposure times
in excess of 6-8 hours may be required for convincing measurements of the fluxes and equivalent widths of such examples.
Given the lower significance of the CIII]$\lambda$1909 detection in GN-108036, we will focus most of the remaining 
discussion on the more secure detection in A383-5.2.  

\section{Modelling the continuum and emission lines of A383-5.2}
The broadband SED of A383-5.2 (Figure 4) exhibits a very strong break between the H$_{160}$-band and the [3.6] and 
[4.5] bands.   Richard et al. (2011) interpret this as a Balmer Break indicative of an evolved (800 Myr) stellar population.   
Alternatively the break could arise if nebular emission from [OIII]+H$\beta$ and H$\alpha$ provide substantial contributions 
within the [3.6] and [4.5] filters, as might be expected if the SED is dominated by a very young stellar population.  Distinguishing 
between these two very different interpretations of the SED is clearly important and can only proceed via direct measurements 
of the rest-frame optical emission lines.  Here we consider whether adding a CIII] line flux constraint, which is sensitive to 
star formation on shorter timescales than that based on the stellar continuum, can clarify the past star formation history of A383-5.2.

\subsection{Method}
We fit the continuum spectral energy distribution and CIII] equivalent width of A383-5.2 using an approach similar to that adopted in 
Stark et al. (2014). We use the latest version of the \citet{Bruzual2003} stellar population synthesis model with the standard 
photoionisation code CLOUDY \citep{Ferland2013} to describe both the stellar and gaseous emission (Gutkin et al., in preparation, 
who follow the prescription of \citealt{Charlot2001}). 

Further details are provided in Stark et al (2014) but, in brief, the main parameters of the photoionised gas are its
interstellar metallicity, $Z$, the typical ionisation parameter of a newly ionised HII region, $U$ (which characterises the ratio of 
ionising-photon to gas densities at the edge of the Str$\rm{\ddot{o}}$mgren sphere), and the dust-to-metal (mass) ratio, 
$\xi_{\mathrm d}$ (which characterizes the depletion of metals on to dust grains). We consider models with C/O (and N/O) 
abundance ratios ranging from 1.0 to 0.05 times the standard values in nearby galaxies [$(\mathrm{C/O})_\odot\approx0.39$ 
and $(\mathrm{N/O})_\odot\approx0.09$] to describe the delayed release of C and N by intermediate-mass stars relative to 
shorter-lived massive stars in young galaxies. We also include attenuation of line and continuum photons by dust in the neutral
 ISM, using the 2-component model of \cite{charlot2000}, as implemented by \citet[][their equations~1--4]{dacunha2008}. This is 
 parameterised in terms of the total $V$-band attenuation optical depth of the dust, $\hat{\tau}_V$, and the fraction $\mu$ of this 
 arising from dust in the diffuse ISM rather than in giant molecular clouds. Accounting for these two dust components is important 
 in differentiating the attenuation of emission-line and stellar continuum photons.

As we will describe in more detail in \S4.2, the need to simultaneously fit the continuum  spectral energy distribution and the CIII] 
equivalent width motivates us to explore a wider range of star formation histories than Stark et al. (2014).   In particular, we 
consider models with two-component star formation histories: a  `starburst' component (represented by a 10\,Myr-old stellar population 
with constant SFR) and an `old' component (represented by a stellar population with constant or exponentially declining SFR with 
age between 10\,Myr and the age of the Universe at the galaxy redshift). We adopt the same stellar metallicity for both components
matching that of the gas-phase.  The relative contribution of the burst component is a free parameter; in the limiting case where the burst 
provides a negligible contribution to the stellar mass fraction, the star formation history approaches the single component star formation 
histories which are commonly used in high redshift SED fitting. Given the young ages at redshifts $z>6$, this simple modeling of 
star formation and chemical enrichment adequately samples the allowed parameter space. We adopt a \citet{chabrier2003} initial 
mass function in all models.

To interpret the combined stellar and nebular emission from A383-5.2, we build a comprehensive grid of models covering wide 
ranges of input parameters. Specifically, we take about 70 model ages between 10\,Myr and the age of the Universe
(at the redshift of the galaxy we are modelling) for the old stellar component; 30 stellar mass fractions of the starburst relative 
to the old components, spaced logarithmically between $10^{-3}$ and unity; metallicities $Z=0.0001$, 0.0002, 0.0005, 0.001,
 0.002, 0.004, 0.008, 0.017, and 0.03; gas ionisation parameters $\log U=-1.0$, $-1.5$, $-2.0$, $-2.5$, $-3.0$, $-3.5$, and $-4.0$; 
 dust-to-metal ratios $\xi_{\mathrm d}=0.1$, 0.3, and 0.5; C/O (and N/O) scaling factors 1.0, 0.85, 0.65, 0.45, 0.25, 0.15, and 0.05; 10 
 attenuation optical depths $\hat{\tau}_V$ between 0 and 1; and, for each $\hat{\tau}_V$, 2 values of $\mu$ (0.3 and 0.9). 
 We adopt a Bayesian approach similar to that of \citet[][their equation~2.10]{pacifici2012} to compute the likelihood of each model given the data.  

\begin{figure}
\begin{center}
\includegraphics[trim=0.0cm 10cm 1cm 5cm, width=0.5\textwidth]{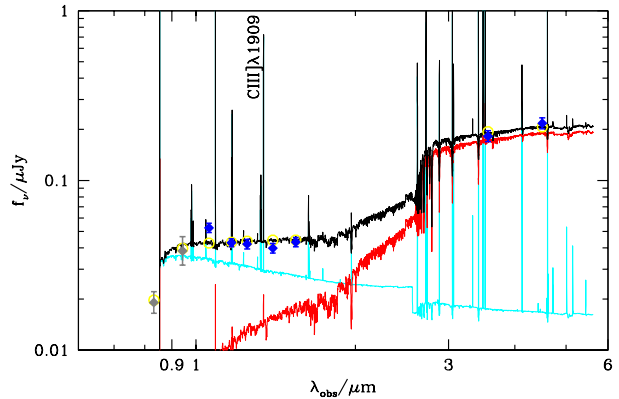}
\caption{SED of A383-5.2 and  population synthesis 
models which provide best fit to the continuum SED and CIII] equivalent width.   
The observed SED is denoted by the diamond data points.     The two grey 
data points at $<$1$\mu$m are not included in the fit because of the uncertainty 
associated with Ly$\alpha$ emission contamination and IGM absorption.    The 
data are best fit by a model with a two component star formation history.   The 
UV continuum and CIII] equivalent width are powered by a recent star formation 
episode (cyan curve), while the optical continuum is dominated by an older 
generation of stars (red curve).   The composite SED is shown in black.     Yellow diamonds 
show predicted broadband fluxes from the best-fitting model.
}
\label{fig:SED}
\end{center}
\end{figure}

\subsection{A383-5.2}

\begin{table}
\begin{tabular}{lr}
\hline 
\multicolumn{2}{c}{Model fit to A383.5.2 }  \\ \hline \hline
                                                $\log{U}$   & $-1.70_{- 0.64}^{+ 0.49}$        \\
  $\log{(M_{\ast,\mathrm{young}}/M_{\ast,\mathrm{tot}})}$   & $-2.99_{- 0.03}^{+ 0.04}$        \\
                                      $\log{(Z/Z_\odot)}$   & $-1.33_{- 0.20}^{+ 0.27}$        \\
                                     $\log(\mathrm{C/O})$   & $-0.58_{- 0.06}^{+ 0.06}$        \\
                                  $\log(\mathrm{age/yr})$   & $ 8.72_{- 0.10}^{+ 0.10}$        \\
                    $\log(M_{\ast,\mathrm{tot}}/M_\odot)$   & $ 9.50_{- 0.10}^{+ 0.10}$        \\
             $\log(\mathrm{SFR}/M_\odot\mathrm{yr}^{-1})$   & $ 0.29_{- 0.08}^{+ 0.08}$        \\
                                           $\hat{\tau}_V$   & $ 0.05_{- 0.05}^{+ 0.05}$        \\
        
\end{tabular}
\caption{\label{obs} Results of fitting procedure for A383-5.2.   Details are provided in \S4.1, and results are discussed in \S4.2.}
\end{table}
The  acceptable parameter fits are shown in Table 1. Including the nebular line constraints in the modelling has provided several 
key advances.  First the dependence of 
emission line equivalent widths on the ionised gas physical conditions allows new constraints 
on the metallicity, relative C/O abundance, and ionisation parameter in star forming galaxies at $z\simeq 6$.   
As can be seen in Table 1, the large equivalent width of CIII] in A383-5.2 
requires models with a very low metallicity (log Z = $-1.33_{- 0.20}^{+ 0.27}$ ), large ionisation parameter 
(log U=$-1.70_{- 0.64}^{+ 0.49}$), and low C/O ratio (log C/O = $-0.58_{- 0.06}^{+ 0.06}$), similar to the population of 
ultra-faint lensed galaxies at $z\simeq 2$ \citep{Stark2014}.   These constraints are almost entirely lost when the 
CIII] equivalent width is not included in the fitting procedure.    
The total stellar mass implied by the models is 3.2$\times$10$^{9}$ M$_\odot$, 
in close agreement with that reported in \citet{Richard2011}.   The current star formation rate is 1.9 M$_\odot$ yr$^{-1}$, broadly consistent 
with that inferred from the Ly$\alpha$ luminosity in \S3.1.1.  

We find that single component star formation histories struggle to reproduce the continuum SED and CIII] 
equivalent width.   Very young ($\lsim 10^{\rm{7}}$ yr), dusty, and 
metal poor models can reproduce the apparent Balmer Break with strong [OIII]+H$\beta$ and H$\alpha$ emission 
dominating the [3.6] and [4.5] fluxes.    But with strong attenuation and substantial reddening, 
the young single component models underpredict the CIII] equivalent width and  overestimate the 
H$_{\rm{160}}$ flux.  Similarly, while an old stellar population produced by a single component star formation 
history can account for the large break between the H$_{\rm{160}}$-band and [3.6], it is unable to reproduce a 
large equivalent width CIII] emission or a blue UV continuum slope supported by the available imaging and spectroscopy.  

We thus find that the observational data  indicate a two-component star formation history (Figure 4).  The nebular lines 
and far UV continuum are dominated by a recent burst, while the rest-optical light seen by the {\it Spitzer}/IRAC filters 
is powered by a somewhat older stellar population.   Both populations contribute roughly equally to the 
emerging near-UV continuum.    The young burst component contributes very little to the total 
stellar mass ($\sim$ 10$^{-3}$) but provides enough ionizing output to match the observed CIII] equivalent width.  

A383-5.2 is certainly not unique in this respect at very high redshift:  \citet{Rodriguez2014} 
have recently argued that a two component star formation history is necessary to explain the combined 
emission  line and continuum constraints on a pair of galaxies at $z=5.07$.   But whether  multi-component 
star formation histories are typical among at $z\gsim 6$ sources is not yet clear.   Very young stellar ages 
($<$10 Myr for constant star formation) have been suggested based on the large equivalent width of optical 
nebular emission lines required to match the {\it Spitzer}/IRAC colours of $z\simeq 7-8$ galaxies 
(e.g. \citealt{Finkelstein2013,Smit2014}).  But it is unlikely that such systems (located 650-800 Myr after 
the Big Bang) have only been forming stars for such a short period.   Indeed the average stellar continuum of 
$z\simeq 8$ galaxies is indicative of older (100 Myr for constant star formation) stars (Dunlop et al. 2013, Labb\'{e} et al. 2013).
Multi-component star formation histories with a recent upturn powering the nebular emission provide a 
natural explanation for this tension.  As larger samples of galaxies with UV metal line and high S/N continuum constraints 
emerge, it will be possible to clarify the nature of early galaxy star formation histories in more detail.  

\section{Reionisation and the Ly$\alpha$ velocity offset}

The attenuation of Ly$\alpha$ emission from galaxies at $z\gsim 6$ provides a valuable probe of reionisation.   
Ground-based spectroscopy has been used by several groups to measure the redshift dependent fraction of UV 
selected galaxies at $z\gsim 6$ with Ly$\alpha$ emission (e.g., \citealt{Stark2010,Schenker2012,Ono2012}).  
As described in \S1, the Ly$\alpha$ emitter fraction appears to drop rapidly over $6<z<8$ (e.g., 
\citealt{Ono2012,Treu2013,Pentericci2014,Schenker2014,Tilvi2014}), as might be expected if the IGM is 
partially neutral  at $z\simeq 7-8$.    

Theoretical efforts are now focused on determining how neutral the IGM must be at $z\simeq 7-8$ 
in order to reproduce the drop in visibility of Ly$\alpha$ emission beyond $z\simeq 6$.
During the late stages of reionisation, the cosmic HII regions surrounding star-forming galaxies are 
expected to be very large. As a result, the impact of IGM damping wing absorption on Ly$\alpha$ emission is likely 
to be minimal in the final phase of reionisation when neutral hydrogen fractions are low.      
Because of this, most attempts to model the drop in the Ly$\alpha$ emitter fraction require 
that the IGM evolves from being fully ionised at $z\simeq 6$ to a neutral fraction by volume as large as 60\% at $z\simeq 7$ 
(e.g., \citealt{McQuinn2007,Mesinger2008,Jensen2013}).

Such a striking evolution in the ionization state of the IGM seems unphysical over such a brief period of 
cosmic history (e.g., \citealt{Sobacchi2014,Robertson2013,Mesinger2014}).   As a result, alternative
explanations have been considered.  If the photo-ionising background drops at $z\gsim 6$, the prevalence of 
self-shielded systems within ionised regions of the IGM would increase rapidly, providing an 
additional source of opacity for Ly$\alpha$ photons \citep{Bolton2013}. Furthermore, if the escape 
fraction of ionising radiation (f$_{\rm{esc}}$) increases by a small amount over  $6<z<7$ ($\Delta$f$_{\rm{esc}}$=0.1), 
the Ly$\alpha$ luminosity may be significantly reduced \citep{Dijkstra2014}. Finally, if the velocity offset between 
Ly$\alpha$ and the galaxy systemic redshift is less than observed at low redshift, a partially neutral IGM would be 
more effective at attenuating  Ly$\alpha$, thereby lowering the neutral fraction required to reproduce the observations
(e.g., \citealt{Dijkstra2011,Schenker2014,Mesinger2014}).

Of the three factors described above, the Ly$\alpha$ velocity offset is the most feasible to constrain via direct 
observations.  If the Ly$\alpha$ velocity offset is similar to that found in bright UV-selected galaxies at $z\simeq 2$ ($\Delta \rm{v_{Ly\alpha}}$=
445 km s$^{-1}$; \citealt{Steidel2010}), the IGM would  likely have to be mostly  neutral at $z\simeq 7$ to be 
consistent with the Ly$\alpha$ observations at the 1$\sigma$ level \citep{Mesinger2014}.  But if the Ly$\alpha$ 
velocity offset is  lower (i.e., $\Delta \rm{v_{Ly\alpha}}$  200 km s$^{-1}$) as is often seen in strong Ly$\alpha$ emitters 
(e.g. \citealt{Tapken2007,McLinden2011,Hashimoto2013}), the downturn in Ly$\alpha$ 
visibility could be produced with an IGM that is still substantially ionised ($\rm{Q{_{HII}}} \lsim 0.6$; 
\citealt{Mesinger2014}) at $z\simeq 7$.  \citet{Schenker2013B} already indicated a possible reduction in the velocity offset with increasing
redshift by comparing observations of [O III] and Ly$\alpha$ in UV-selected galaxies at $z\simeq 3.5$ with similar samples at $z\simeq 2$. 
Albeit with a small sample of 9 sources, their velocity offsets were typically $\simeq$150 km s$^{-1}$.

However, at $z\gsim 4$, such rest-frame optical emission lines are not detectable from the ground.   
Thus, since the UV metal lines probe the galaxy systemic redshift (e.g., \citealt{Erb2010}, Stark et al. 2014), they 
provide the only immediate means of determining the distribution of $\Delta \rm{v_{Ly\alpha}}$ at $z\simeq 6$.  
Indeed, our detection of CIII]$\lambda$1909 in A383-5.2 provides the first constraint on the 
Ly$\alpha$ velocity offset in a $z\gsim 5$ galaxy, and reveals an offset of only $\rm{120~km~s^{-1}}$ (Figure 5). 
Such a low Ly$\alpha$ velocity offset is consistent with an extension of the redshift-dependent trend found in \citet{Schenker2013B}, 
likely reflecting the  general increase in the Ly$\alpha$ equivalent width over $3<z<6$ seen for similarly-selected UV sources \citep{Stark2011}.

The detection of CIII]$\lambda$1909 in Figure 3, although less significant, implies that the Ly$\alpha$ offset of GN-108056 
is possibly even smaller.  While A383-5.2 is observed when the Universe is highly ionized, GN-108056  is likely observed in a 
partially neutral IGM.   At $z\gsim 7$, the IGM damping wing absorption will attenuate Ly$\alpha$ most strongly in systems with low 
velocity offsets.   Focusing our UV metal line follow-up on galaxies with known Ly$\alpha$ detections should bias us toward 
galaxies with significant transmission through the surrounding IGM.   In principle, this should cause us to select the subset of 
Ly$\alpha$ emitters with large velocity offsets.    The fact that we find  little to no velocity offset for GN-108036, one of the few known 
Ly$\alpha$ emitters at $z\simeq 7$, might reflect the absence of large velocity offsets in the reionization era.  

If  Ly$\alpha$ velocity offsets as small as A383-5.2 and GN-108036 are common at $z\gsim 6$, the IGM evolution required 
to reproduce the downturn in the Ly$\alpha$ fraction would be  less rapid than 
previous estimates using large Ly$\alpha$ velocity offsets , providing a more physically reasonable reionisation history (see \citealt{Robertson2010,Robertson2013,Mesinger2014}).    As the 
number of bright $z\gsim 6$ galaxies with Ly$\alpha$ detections increases, it will be
feasible to assemble a moderate sample of $z\simeq 6$ galaxies with Ly$\alpha$ and UV 
metal line detections, providing the first measurement of the distribution of Ly$\alpha$ velocity offsets 
in reionisation-era galaxies and thereby removing one of the key systematic uncertainties in mapping Ly$\alpha$ 
evolution to a reionisation history.

\begin{figure}
\begin{center}
\includegraphics[trim=3.0cm 13cm 7cm 5cm,width=0.48\textwidth]{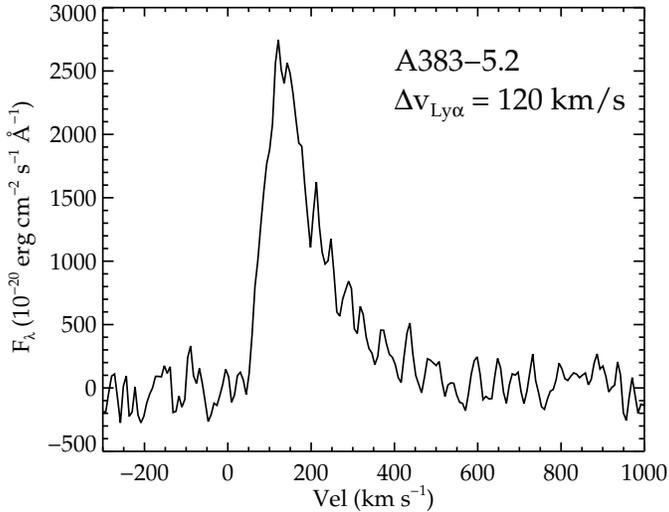}
\caption{Velocity profile of Ly$\alpha$ emission in the $z=6.027$ galaxy A383-5.2.   Ly$\alpha$ is 
shifted to the rest-frame using the systemic redshift provided by CIII]$\lambda$1909.   The peak flux 
of Ly$\alpha$ emission is shifted by $\Delta v$=120 km s$^{-1}$ from the systemic redshift.}
\label{fig:profile}
\end{center}
\end{figure}

\section{Discussion and Summary}

Considerable effort has been placed in spectroscopic study of  $z\gsim 6$ galaxies. Yet, in spite of 
significant allocations of telescope time to several research teams, few spectroscopic redshifts have been confirmed 
beyond $z\simeq 7$ due to the absence of Ly$\alpha$. Although the physical explanation for
this suppression is of interest in its own right in terms of constraining the reionisation history, it is essential 
to locate lines other than Ly$\alpha$ in distant sources for many reasons. Firstly, the Ly$\alpha$ emitter fraction test 
is currently applied to a photometrically-selected sample. If no line is detected, there is an obvious ambiguity 
in interpretation between a suppressed Ly$\alpha$ at the redshift of interest and a foreground object.
Clearly securing the redshift from an additional line breaks this degeneracy.
Secondly, as discussed in \S5, the interpretation of the redshift-dependent Ly$\alpha$ fraction 
in terms of the evolving neutral fraction depends on the velocity offset of the line with respect
to the systemic redshift which, at $z>6$, can only be determined with current facilities
using rest-frame UV nebular lines. Finally, Stark et al (2014) demonstrate, via detailed modeling,
how measures of rest-frame UV lines such as NIV], OIII], CIV, Si III] and CIII] can provide unique
information on important physical properties of galaxies during the reionisation era, including 
the ionisation parameters, metallicities and star formation histories which remain degenerate
when interpreting SEDs based on broad-band photometry alone.

In this paper we have demonstrated it is feasible to detect the CIII] 1908 \AA\
doublet in $z>6$ star-forming galaxies. The two sources we discuss are very different
and illustrate the challenges even with state-of-the-art near-infrared spectrographs
on the largest ground-based telescopes. A363-5.2 is a gravitationally-lensed 
galaxy at a redshift $z_{\rm{Ly\alpha}}=$6.029. It is bright ($J_{\rm 125}=25.2$), highly magnified ($\times$7.4) 
and a convincing CIII] 1909 detection has been secured in only a 3.5 hour exposure. The source is typical
of the brightest sources that have been located in lensing surveys of
foreground clusters (e.g. the CLASH program). GN-108036 is more typical of the bright 
sources found in deep blank field surveys. Although not  fainter ($J_{\rm 140}=25.2$), as it is 
unlensed, it is more luminous and even with a 4.2 hour exposure, the detected CIII] is much
weaker. 

The robust detection of CIII] in A383-5.2 has enabled us to measure a reduced velocity
offset of Ly$\alpha$ in a $z>6$ source as well as to break degeneracies of
interpretation in the SED. Specifically, including the strength of CIII] in our modeling
fits to the spectral energy distribution allows us to determine a low metallicity
(log Z=$-1.33$), a large ionisation parameter (log U=$-1.70$) and a low C/O ratio.
More importantly, however, we argue the tension between the strength of CIII] (which
is sensitive to star formation on 10 Myr timescales) and the measured Balmer break and rest-frame
UV continuum slope (indicative of a more mature population) can be reconciled if
the star formation has been erratic with a recent burst of activity contributing significantly
to the nebular spectrum. If this behavior is typical of $z>6$ star-forming galaxies, it
may explain the prominent rest-frame optical nebular lines inferred in recent studies
(e.g. Smit et al 2013).

As discussed in \S1, the ultimate goal would be a redshift survey based on rest-frame UV lines other
than Ly$\alpha$, particularly in securing more precise measures of the Ly$\alpha$ fraction. However, 
we have considered it is prudent in the first instance to base our searches using
those few targets for which Ly$\alpha$ has already been detected. This may, admittedly,
lead to some biases in interpretation if, for example, the sources have untypical Ly$\alpha$
fluxes for some reason. As one progresses to higher redshift, the flux ratio of CIII] to Ly$\alpha$
should  increase since Ly$\alpha$ is increasingly suppressed and CIII] is
more prominent in metal-poor systems. Figure 6 gives no strong evidence of a
shift in this direction compared to the distribution seen at lower redshift by Stark et al (2014)
but the samples are small and uncertainties remain large.

\begin{figure}
\begin{center}
\includegraphics[width=0.48\textwidth]{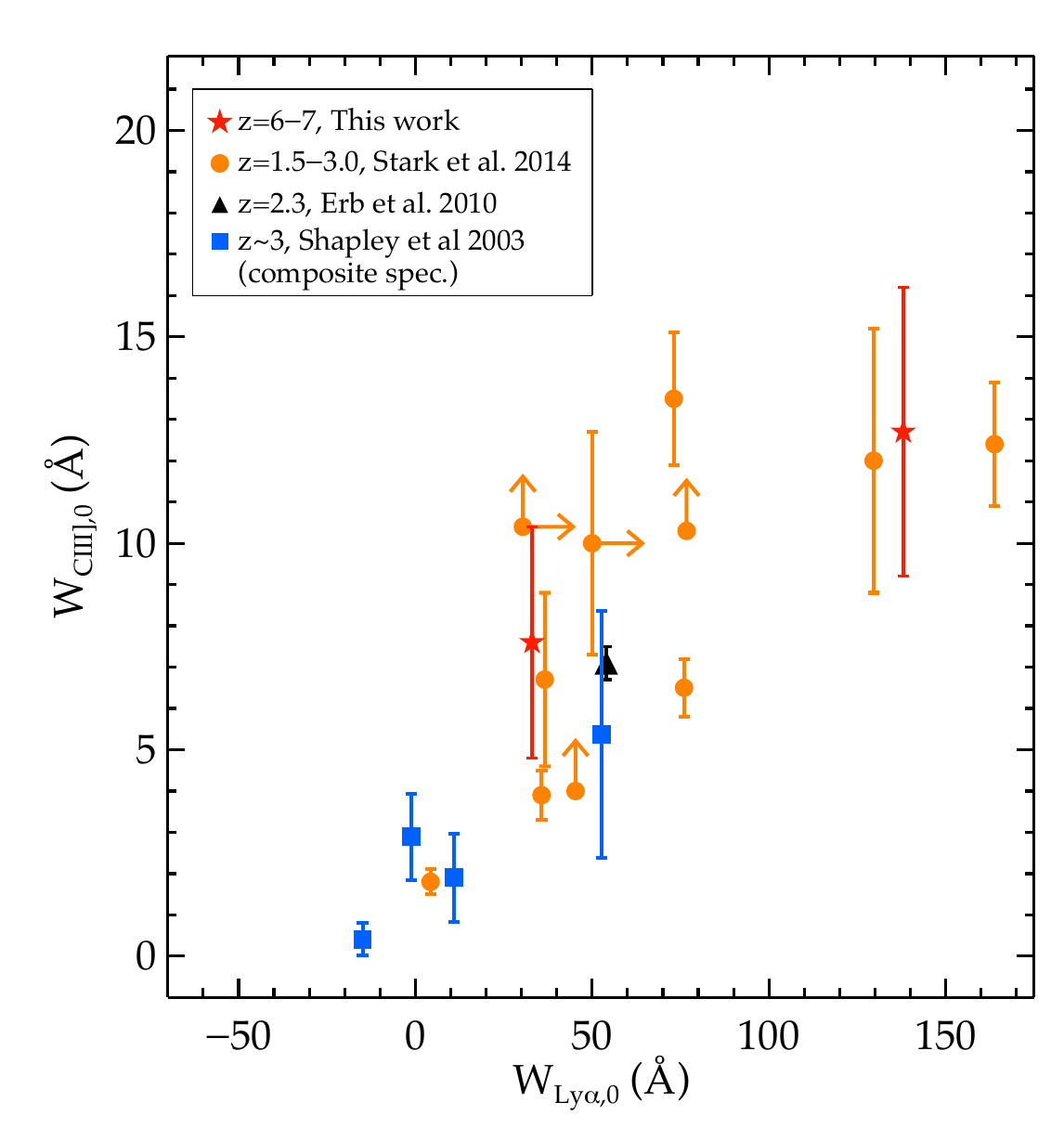}
\caption{ A comparison of the rest-frame equivalent
widths of CIII] 1909 \AA\ and Ly$\alpha$ from the study of $z\simeq 2$ lensed sources by  
Stark et al (2014, orange points), Erb et al (2010) and the z$\simeq 3$ Shapley et al (2003)
composite alongside the two new CIII] detections beyond $z\simeq 6$ (this paper, red stars).
}
\label{fig:profile}
\end{center}
\end{figure}

Despite the observational challenges, our results confirm the feasibility of
studying bright galaxies in the reionisation era through rest-frame UV lines
other than Ly$\alpha$. 

\section*{Acknowledgments}
We thank Dawn Erb, Martin Haehnelt, Juna Kollmeier, Andrei Mesinger, 
and Alice Shapley for enlightening conversations.   DPS acknowledges support from the 
National Science Foundation  through the grant AST-1410155.   JR acknowledges support 
from the European Research Council (ERC) starting grant CALENDS and the Marie Curie Career Integration Grant 294074.   
SC, JG and AW acknowledge support from the ERC via an Advanced Grant under grant agreement no. 321323 -- NEOGAL.  
The results are partially based on observations made with ESO telescopes at 
the La Silla Paranal Observatory under programme 092.A-0630 and the W.M. Keck
Observatory. This work was partially supported  by a NASA Keck PI Data Award, administered by the 
NASA Exoplanet Science Institute. Data presented herein were obtained at the W. M. Keck Observatory 
from telescope time allocated to the National Aeronautics and Space Administration through the agency's 
scientific partnership with the California Institute of Technology and the University of California. 
The Observatory was made possible by the generous financial support of the W. M. Keck Foundation.
The authors acknowledge the very significant cultural role that the
summit of Mauna Kea has always had within the indigenous Hawaiian community.
We are most fortunate to have the opportunity to conduct observations from this mountain.

\bibliographystyle{mn2e}
\bibliography{references}


\label{lastpage}

\end{document}